\documentclass[11pt,a4paper]{JHEP3}
\pdfoutput=1
\usepackage{epsfig}
\usepackage{amsmath}

\title{Quarter-BPS states in orbifold sigma models with ADE singularities}

\author{Kenny Wong\\Department of Applied Mathematics and Theoretical Physics, \\Centre for Mathematical Sciences, \\University of Cambridge, \\Cambridge, CB3 0WA, UK\\{\tt k.wong@damtp.cam.ac.uk}}

\abstract{We study the elliptic genera of two-dimensional orbifold CFTs, where the orbifolding procedure introduces du Val surface singularities on the target space. The $\mathcal N = 4$ character decompositions of the elliptic genus contributions from the twisted sectors at the singularities obey a consistent scaling property, and contain information about the arrangement of exceptional rational curves in the resolution. We also discuss how these twisted sector elliptic genera are related to twining genera and Hodge elliptic genera for sigma models with K3 target space.}

\begin{document}
\maketitle
\flushbottom

\section{Introduction}

\paragraph{}
Du Val singularities are a class of surface singularities, realised by quotienting $\mathbf C^2$ by a discrete subgroup  $G$ of $ SU(2)$. The exceptional divisor in the minimal resolution of a du Val singularity is a chain of rational curves, whose intersection matrix is the Cartan matrix of an ADE root system. A supersymmetric sigma model with target space $\mathbf C^2 / G$ can be constructed by orbifolding the free theory on $\mathbf C^2$ by the discrete group $G$, and the counting of  half-BPS multiplets\footnote{In this article, the theories have two-dimensional $\mathcal N = (4,4)$ superconformal symmetry. A ``half-BPS'' multiplet is defined to be a tensor product of a left-moving short multiplet with a right-moving short multiplet, whereas a ``quarter-BPS'' multiplet is defined to be a tensor product of a left-moving long multiplet with a right-moving short multiplet. Thus, the only multiplets with non-zero Witten index are the half-BPS multiplets.} in the twisted sectors of these orbifolds provides a celebrated physical explanation \cite{wittenmckay,wittenmckay2} for the McKay correspondence \cite{mckay} that relates the number of conjugacy classes in the group $G$ to the number of rational curves in the minimal resolution of $\mathbf C^2 / G$, or equivalently, to the number of nodes in the associated ADE Dynkin diagram (see for instance the reviews \cite{reidreview,otherreview}).

\paragraph{}
In this article, we study the spectra of \emph{quarter-BPS} multiplets in the twisted sectors of these $\mathbf C^2 / G$ orbifolds. These spectra are encoded in the twisted sector contributions to the elliptic genus. We  observe that the multiplicities of quarter-BPS multiplets obey a consistent pattern:  as the number of rational curves in the exceptional divisor increases, the multiplicities of sufficiently low-lying quarter-BPS multiplets grow in proportional to a sequence,
\begin{eqnarray*}
 4, \ \ 20, \ \ 66, \ \ 194, \ \ 492, \ \ 1178, \ \ 2606, \ \ 5530 ,  \ \ \dots 
\end{eqnarray*}
whose origin, as far as the author is aware, is a mystery.

\paragraph{}
The quarter-BPS spectrum has a further interesting property in that there is a natural way to pair the conjugacy classes in $G$ with the  exceptional rational curves in the resolution, and the multipliticities of quarter-BPS multiplets in the twisted sectors associated to a given conjugacy class are fully determined by how the corresponding rational curve is positioned in relation to the other rational curves in the resolution. These patterns will be described in Section 2.

\paragraph{}
These twisted sector genera are common to all orbifold CFTs with du Val type singularities, the $\mathbf C^2 /G$ orbifold being only one example. For this reason, they are very natural quantities to study. Most notably, du Val singularities appear in certain orbifold constructions of sigma models with K3 target space, and the twisted sector contributions from the ADE singular points serve as building blocks for the full elliptic genera of these K3 sigma models. The  recent discovery of Mathieu moonshine \cite{moonshine1, moonshine2, moonshine3} showed that the multiplicities of quarter-BPS multiplets in K3 sigma models themselves obey a remarkable numerical property in that they coincide with  dimensions of  non-trivial representations of the sporadic group $M_{24}$. Since certain K3 geometries can be obtained by orbifolding other K3 geometries, and since the orbifolding procedure introduces du Val singularities, one would expect that the elliptic genus contributions from twisted sectors around du Val singularities, which obey the curious numerical properties described above, should be related to the twined K3 elliptic genera  of Mathieu moonshine. This intricate network of relationships, along with other connections to K3 sigma models, is explained in Section 3.

\section{The $\mathbf C^2 / G$ orbifold}

\paragraph{}
Since $\mathbf{ C}^2$ is a hyperk\"ahler manifold,  there is a natural two-dimensional $\mathcal N = (4,4)$ superconformal  sigma model with $\mathbf{C}^2$ target space. It consists of a pair of free complex bosons parametrising  $\mathbf {C}^2$, and pairs of free left- and right-moving complex fermions taking values in the tangent bundle of $\mathbf C^2$. The central charge is $c = \bar c = 6$. The natural $SU(2)$ group acting on the coordinates of $\mathbf {C}^2 $ preserves the hyperk\"ahler structure, so any orbifold of the sigma model by a subgroup $G \subset SU(2)$ inherits the $\mathcal N = (4,4)$ superconformal symmetry.

\paragraph{}
When $G$ is a finite subgroup of $SU(2)$, the space $\mathbf {C}^2 / G$ has a du Val surface singularity at the origin, whose preimage in the minimal resolution consists of a collection of rational curves with intersection matrix given by an ADE Dynkin diagram:
\begin{itemize}
\item The $A_k$ singularity: $G$ is the cyclic group of order $k + 1$.
\item The $D_k$ singularity: $G$ is the binary dihedral group of order $4k - 8$.
\item The $E_6$ singularity: $G$ is the binary tetrahedral group of order 24.
\item The $E_7$ singularity: $G$ is the binary octohedral group of order 48.
\item The $E_8$ singularity: $G$ is the binary icosahedral group of order 120.
\end{itemize}
Explicit generators for these discrete subgroups are provided in the Appendix.

\paragraph{}
Our quantity of interest  is the elliptic genus of the theory. The elliptic genus \cite{witteneg,othereg} is a partition function defined as the following trace over Ramond-Ramond states:
\begin{eqnarray*}
Z(q, \bar q, y) = {\rm Tr}_{\rm RR} (-1)^F q^{L_0} \bar q^{\bar L_0} y^{J_0}.
\end{eqnarray*}
Here, $F$ is the fermion number operator, $L_0$ and $\bar L_0$ are the left- and right-moving Virasoro generators, and $J_0$ is a  left-moving R-symmetry generator. The elliptic genus contains information about the degeneracies of half-BPS and quarter-BPS multiplets in the theory.

\paragraph{}
Because the orbifold theory\footnote{Note that, although $\mathbf{C}^2 / G$ is singular as an algebraic variety, the CFT is not singular; there is a non-trivial B-field flux through the exceptional $\mathbf P^1$'s in the CFT. This orbifold CFT is different from the model of an ADE singularity in \cite{ade4,anotherliouville}, constructed as a discrete quotient of an  $SL(2,\mathbf R)/U(1)$ coset tensored with an $\mathcal N = 2$ minimal model, where the B-field fluxes through the exceptional $\mathbf P^1$'s are tuned to zero.} on $\mathbf C^2 / G$ is obtained from the free theory on $\mathbf C^2$ by introducing twisted sectors, and by projecting onto states invariant under $G$, the elliptic genus for the orbifold can be written as a sum over twisted and twined elliptic genera of the free theory,
\begin{eqnarray*}
Z_{\mathbf C^2 / G}(q,\bar q, y) = \frac 1 {|G|} \sum_{g \in G} \sum_{h \in G} Z_{\mathbf C^2,(g,h)}(q, \bar q, y),
\end{eqnarray*}
where
$$  Z_{\mathbf C^2,(g,h)}(q,\bar q, y) := {\rm Tr}_{ g{\rm -twisted, RR}} \  h  (-1)^F  q^{L_0} \bar q^{\bar L_0} y^{J_0}. $$

\paragraph{}
The untwisted sector contribution to $Z_{\mathbf C^2 / G}$ (i.e. the contribution with $g$ equal to the identity element) includes a sum over a continuous spectrum of scattering states. While there are ways of regularising this sum (see \cite{sungjay} for example), this untwisted sector is not the main focus of this article, and will play no further part in our discussion.

\paragraph{}
We will instead focus on the contribution to $Z_{\mathbf C^2 / G}(q, \bar q, y)$ from the twisted sectors,
\begin{eqnarray*}Z_{\mathbf C^2 / G,{\rm twisted}}(q, y) : = \frac 1 {|G|} \sum_{g \in G \backslash \{ e \} } \sum_{h \in G} Z_{\mathbf C^2,(g,h)}(q, y), \end{eqnarray*}
which is a well-defined sum over a discrete spectrum of states. In the second half of this article, we will explain how the quantity $Z_{\mathbf C^2/G, {\rm twisted}}$ can be viewed as a universal quantity valid for any orbifold sigma model where the orbifolding introduces a du Val singularity of the same ADE type.

\paragraph{}
As already mentioned, the elliptic genus encodes information about the half-BPS and quarter-BPS spectrum of the sigma model. The multiplets of the $\mathcal N = 4$ superconformal algebra at central charge $c = 6$ are as follows:
\begin{itemize}
\item A short multiplet whose highest weight state has eigenvalues $(L_0, J_0) = (h,j) =( \frac 1 4, 0)$. This multiplet has Witten index $1$.
\item A short multiplet whose highest weight state has eigenvalues $(L_0, J_0) = (h,j)=( \frac 1 4 , \frac 1 2 )$. This multiplet has Witten index $-2$.
\item A sequence of long multiplets, labelled by integers $n \in \{1,2,3, \dots \}$, whose highest  weight states have eigenvalues $(L_0, J_0) = (h,j) =( \frac 1 4 + n , \frac 1 2 )$. These multiplets all have zero Witten index.
\end{itemize}

\paragraph{}
From the perspective of the right-movers, the elliptic genus is merely a Witten index. So the elliptic genus only receives contributions from short multiplets in the right-moving sector. In particular, it is independent of $\bar q$, at least in the twisted sectors (where there are no issues with regularisation), which explains why the $\bar q$ variable is dropped from our notation. However, the elliptic genus receives contributions from both short and long multiplets in the left-moving sector. Defining the character of the left-moving $\mathcal N = 4$ multiplet with highest weight $(h,j)$ as 
\begin{eqnarray*}
{\rm ch}_{h,j} (q,y) = {\rm Tr}_{\mathcal H_{(h,j)}}(-1)^Fq^{L_0} y^{J_0},
\end{eqnarray*}
one can decompose the twisted sector contribution to the elliptic genus of $\mathbf C^2 / G$ as a linear combination of these characters,
\begin{eqnarray*}
Z_{\mathbf C^2 / G,{\rm twisted}} (q, y)= a_{\frac 1 4 , 0 } {\rm ch}_{\frac 1 4 , 0} (q,y)+  a_{\frac 1 4 , \frac 1 2 } {\rm ch}_{\frac 1 4 , \frac 1 2} (q,y)+ \sum_{n = 1}^\infty a_{\frac 1 4 + n, \frac 1 2 } {\rm ch}_{\frac 1 4 + n, \frac 1 2} (q,y).
\end{eqnarray*}
Explicit expressions for the characters ${\rm ch}_{h,j} (q,y) $ have been computed in \cite{characters1, characters2}, and are reproduced in the Appendix for convenience.

\paragraph{}
The full spectrum of twisted sector states decomposes into a direct sum of tensor products of left-moving and right-moving $\mathcal N = 4$ multiplets,
$$ \mathcal H_{\mathbf C^2 / G, {\rm twisted}} = \bigoplus_{(h,j),(\bar h, \bar j)}  c_{(h,j),(\bar h, \bar j)} \mathcal H_{(h,j)}^{\rm LM} \otimes \mathcal H_{(\bar h, \bar j)}^{\rm RM},$$
where the integer $ c_{(h,j),(\bar h, \bar j)}$ denotes the multiplicity with which the tensor product of the left-moving $(h,j)$-multiplet and the right-moving $(\bar h , \bar j)$-multiplet appears in the twisted sector spectrum.

\paragraph{}
Each coefficient $a_{(h,j)}$ can then be interpreted as the sum of the Witten indices\footnote{In fact, for twisted sectors on $\mathbf C^2 / G$, we simply have $a_{(h,j)} = c_{(h,j), (\frac 1 4, 0)}$. This can be verified by considering refined versions of the elliptic genus, such as the Hodge elliptic genus introduced in \cite{arnav1, arnav2, nathan}. However, it is not clear that this simplification must occur generally for ADE-type singularities in orbifolds of arbitrary  spaces, which will be our focus in the next section, so we do not emphasise this point.} of all right-moving short multiplets paired with the left-moving $(h,j)$-multiplet  in the twisted sector spectrum:
\begin{eqnarray*}
a_{(h,j)} = c_{(h,j), (\frac 1 4, 0)} - 2 c_{(h,j), (\frac 1 4 , \frac 1 2 )}.
\end{eqnarray*} 
Thus the $a_{(h,j)}$ coefficients can be viewed as signed counts of half-BPS (short-short) multiplets and quarter-BPS (long-short) multiplets in the twisted sectors.

\paragraph{}
The elliptic genus and the character decompositions can computed by standard techniques, as described in the Appendix. Here, we proceed directly to the results, beginning with the $A_k$ type singularities:

\begin{center}
\begin{tabular}{c|cc|ccccccccc}
&$ a_{\frac 1 4,0} $ &  $a_{\frac 1 4, \frac 1 2}$ & $a_{\frac 5 4, \frac 1 2}$ & $a_{\frac 9 4, \frac 1 2}$ & $a_{\frac {13} 4, \frac 1 2}$ & $a_{\frac {17} 4, \frac 1 2}$ & $a_{\frac {21} 4, \frac 1 2}$ & $a_{\frac {25} 4, \frac 1 2}$ & $a_{\frac {29} 4, \frac 1 2}$ & $a_{\frac {33} 4, \frac 1 2}$ & $ \dots $\\
\hline
$A_1$ & 1 & 0 & 6 & 28 & 98 & 282 & 728 & 1734 & 3864 & 8182 & $ \dots $\\
$A_2$ & 2 & 0 & 10 & 52 & 170 & 506 & 1290 & 3090 & 6854 & 14570 & $ \dots $ \\
$A_3$ & 3 & 0 & 14 & 72 & 242 & 710 & 1812 & 4346 & 9644 & 20478  & $ \dots $\\
$A_4$ & 4 & 0 & 18 & 92 & 308 & 912 & 2318 & 5566 & 12336 & 26220 & $ \dots $ \\
$A_5$ & 5 & 0 & 22 & 112 & 374 & 1106 & 2820 & 6762 & 14996 & 31862 & $ \dots $ \\
$A_6$ & 6 & 0 & 26 & 132 & 440 & 1300 & 3312 & 7952 & 17624 & 37458 & $ \dots $ \\
$A_7$ & 7 & 0 & 30 & 152 & 506 & 1494 & 3804 & 9130 & 20244 & 43014 & $ \dots $ \\
$A_8$ & 8 & 0 & 34 & 172 & 572 & 1688 & 4296 & 10308 & 22850 & 48560 & $ \dots $ \\
$A_9$ & 9 & 0 & 38 & 192 & 638 & 1882 & 4788 & 11486 & 25456 & 54090 & $ \dots $ \\
$\vdots$ & $\vdots$ & $\vdots$ & $\vdots$ & $\vdots$ & $\vdots$ & $\vdots$ & $\vdots$ & $\vdots$ & $\vdots$ & $\vdots$ 

\end{tabular}
\end{center}

\paragraph{}
The half-BPS spectrum, given in the two columns on the left, is well-known and well-understood \cite{wittenmckay, wittenmckay2}. The quantity $a_{\frac 1 4, 0} - 2a_{\frac 1 4, \frac 1 2}$ is the Witten index contribution from the twisted sectors, which is equal to the number of exceptional rational curves in the minimal resolution of the $A_k$ singularity, consistent with the expectation that the Witten index contribution from the twisted sectors should be the difference between the Euler characteristics of the resolved and unresolved geometries\footnote{Since the resolution of an ADE singularity is a continuous deformation of the CFT \cite{ade4,ade1}, the full Witten index of the theory is equal to the Euler characteristic of the resolved geometry. Meanwhile, the Witten index of the untwisted sector, computed from $G$-invariant states on the space before quotienting, is the Euler characteristic of the unresolved quotient space. The Witten index of the twisted sectors is their difference. Of course, this description is problematic for $\mathbf C^2 / G$ because of the continuous spectrum of untwisted sector states, which needs to be regularised; however, the same argument applies for the compact geometries that we shall consider in Section 3, where there are no issues with regularisation. The fact that the Witten index contributions come from $a_{\frac 1 4, 0}$ rather than from $a_{\frac 1 4, \frac 1 2}$ can be attributed to the observation that the $(h,j) = (\frac 1 4, 0)$ multiplet contributes $+1$ to the Hirzebruch $\chi_y$ genus, consistent with the fact that the divisor class of an exceptional $\mathbf P^1$ is contained in $H^{1,1}$.}. Moreover, the Witten index of the twisted sector is also equal to the number of non-identity conjugacy classes in $G$ (see the Appendix for details). This correspondence between conjugacy classes in $G$ and exceptional rational curves is a manifestation of the McKay correspondence \cite{wittenmckay, wittenmckay2, mckay}.

\paragraph{}
Our main interest in this article, however, is in the coefficients $a_{\frac 1 4 + n, \frac 1 2 }$ for $n \geq 1$, which provide signed counts of the quarter-BPS multiplets in the twisted sectors. A pattern emerges when we compute the \emph{differences} between the signed quarter-BPS counts between successive values of $k$:

\begin{center}
\begin{tabular}{c|cc|ccccccccc}
&$ a_{\frac 1 4,0} $ &  $a_{\frac 1 4, \frac 1 2}$ & $a_{\frac 5 4, \frac 1 2}$ & $a_{\frac 9 4, \frac 1 2}$ & $a_{\frac {13} 4, \frac 1 2}$ & $a_{\frac {17} 4, \frac 1 2}$ & $a_{\frac {21} 4, \frac 1 2}$ & $a_{\frac {25} 4, \frac 1 2}$ & $a_{\frac {29} 4, \frac 1 2}$ & $a_{\frac {33} 4, \frac 1 2}$ & $ \dots $ \\
\hline
$A_0 \to A_1$ & 1 & 0 & 6 & 28 & 98 & 282 & 728 & 1734 & 3864 & 8182  & $ \dots $\\
$A_1 \to A_2$ & 1 & 0 & \textbf{4} & 24 & 72 & 224 & 562 & 1356 & 2990 & 6388 & $ \dots $\\
$A_2 \to A_3$ & 1 & 0 & \textbf{4} & \textbf{20} & 72 & 204 & 522 & 1256 & 2790 & 5908 & $ \dots $ \\
$A_3 \to A_4$ & 1 & 0 & \textbf{4} & \textbf{20} & \textbf{66}& 202 & 506 & 1220 & 2692 & 5742 & $ \dots $\\
$A_4 \to A_5$ & 1 & 0 & \textbf{4} & \textbf{20} & \textbf{66} & \textbf{194} & 502 & 1196 & 2660 & 5642 & $ \dots $ \\
$A_5 \to A_6$ & 1 & 0 & \textbf{4} & \textbf{20} & \textbf{66} & \textbf{194} & \textbf{492} & 1190 & 2628 & 5596 & $ \dots $\\
$A_6 \to A_7$ & 1 & 0 & \textbf{4} & \textbf{20} & \textbf{66} & \textbf{194} & \textbf{492} & \textbf{1178} & 2620 & 5556 & $ \dots $\\
$A_7 \to A_8$ & 1 & 0 & \textbf{4} & \textbf{20} & \textbf{66} & \textbf{194} & \textbf{492} & \textbf{1178} & \textbf{2606} & 5546 & $ \dots $\\
$A_8 \to A_9$ & 1 & 0 & \textbf{4} & \textbf{20} & \textbf{66} & \textbf{194} & \textbf{492} & \textbf{1178} & \textbf{2606} & \textbf{5530} & $ \dots $\\
$\vdots$ & $\vdots$ & $\vdots$ & $\vdots$ & $\vdots$ & $\vdots$ & $\vdots$ & $\vdots$ & $\vdots$ & $\vdots$ & $\vdots$ 

\end{tabular}
\end{center}

\paragraph{}
The signed counts of sufficiently low-lying quarter-BPS multiplets increase by a constant increment every time the number of nodes $k$ in the $A_k$ Dynkin diagram increases. To be more precise, the increment in $a_{\frac 1 4 + n, \frac 1 2}$ from $A_k$ to $A_{k+1}$ stabilises when $k \geq n$. (The low-lying multiplets to which this pattern applies are highlighted for emphasis.)  The regular increment is given by the sequence:

\begin{center}
\begin{tabular}{c|cc|ccccccccc}
&$ a_{\frac 1 4,0} $ &  $a_{\frac 1 4, \frac 1 2}$ & $a_{\frac 5 4, \frac 1 2}$ & $a_{\frac 9 4, \frac 1 2}$ & $a_{\frac {13} 4, \frac 1 2}$ & $a_{\frac {17} 4, \frac 1 2}$ & $a_{\frac {21} 4, \frac 1 2}$ & $a_{\frac {25} 4, \frac 1 2}$ & $a_{\frac {29} 4, \frac 1 2}$ & $a_{\frac {33} 4, \frac 1 2}$ & \dots \\
\hline
$A_\infty\to A_{\infty+1}$ & 1 & 0 & 4 & 20 & 66 & 194 & 492 & 1178 & 2606 & 5530 & \dots \\
\end{tabular}
\end{center}

\paragraph{}
Now, let us examine the same data for the $D_k$ singularities:

\begin{center}
\begin{tabular}{c|cc|ccccccccc}
&$ a_{\frac 1 4,0} $ &  $a_{\frac 1 4, \frac 1 2}$ & $a_{\frac 5 4, \frac 1 2}$ & $a_{\frac 9 4, \frac 1 2}$ & $a_{\frac {13} 4, \frac 1 2}$ & $a_{\frac {17} 4, \frac 1 2}$ & $a_{\frac {21} 4, \frac 1 2}$ & $a_{\frac {25} 4, \frac 1 2}$ & $a_{\frac {29} 4, \frac 1 2}$ & $a_{\frac {33} 4, \frac 1 2}$ & $ \dots $\\
\hline
$D_4$ & 4 & 0 & 18 & 94 & 314 & 924 & 2354 & 5652 & 12534 & 26626& $ \dots $ \\
$D_5$ & 5 & 0 & 22 & 114 & 380 & 1122 & 2858 & 6860 & 15210 & 32318 & $ \dots $\\
$D_6$ & 6 & 0 & 26 & 134 & 446 & 1316 & 3350 & 8044 & 17834 & 37894 & $ \dots $ \\
$D_7$ & 7 & 0 & 30 & 154 & 512 & 1510 & 3842 & 9222 & 20440 & 43432 & $ \dots $\\
$D_8$ & 8 & 0 & 34 & 174 & 578 & 1704 & 4334 & 10400 & 23046 & 48962 & $ \dots $ \\
$\vdots$ & $\vdots$ & $\vdots$ & $\vdots$ & $\vdots$ & $\vdots$ & $\vdots$ & $\vdots$ & $\vdots$ & $\vdots$ & $\vdots$ 

\end{tabular}
\end{center}

\paragraph{}
Once again, a pattern emerges upon taking differences between the signed quarter-BPS counts between successive $D_k$ singularities:

\begin{center}
\begin{tabular}{c|cc|ccccccccc}
&$ a_{\frac 1 4,0} $ &  $a_{\frac 1 4, \frac 1 2}$ & $a_{\frac 5 4, \frac 1 2}$ & $a_{\frac 9 4, \frac 1 2}$ & $a_{\frac {13} 4, \frac 1 2}$ & $a_{\frac {17} 4, \frac 1 2}$ & $a_{\frac {21} 4, \frac 1 2}$ & $a_{\frac {25} 4, \frac 1 2}$ & $a_{\frac {29} 4, \frac 1 2}$ & $a_{\frac {33} 4, \frac 1 2}$ & $ \dots $ \\
\hline
$D_4 \to D_5$ & 1 & 0 & \textbf{4} & \textbf{20} & \textbf{66} & 198 & 504 & 1208 & 2676 & 5692 & $ \dots $ \\
$D_5 \to D_6$ & 1 & 0 & \textbf{4} & \textbf{20} & \textbf{66} & \textbf{194} & \textbf{492} & 1184 & 2624 & 5576 & $ \dots $ \\
$D_6 \to D_7$ & 1 & 0 & \textbf{4} & \textbf{20} & \textbf{66} & \textbf{194} & \textbf{492} & \textbf{1178} & \textbf{2606} & 5538 & $ \dots $ \\
$D_7 \to D_8$ & 1 & 0 & \textbf{4} & \textbf{20} & \textbf{66} & \textbf{194} & \textbf{492} & \textbf{1178} & \textbf{2606} & \textbf{5530} & $ \dots $ \\
$\vdots$ & $\vdots$ & $\vdots$ & $\vdots$ & $\vdots$ & $\vdots$ & $\vdots$ & $\vdots$ & $\vdots$ & $\vdots$ & $\vdots$ 

\end{tabular}
\end{center}

\paragraph{}
The increments stabilise to the same sequence as for the $A_k$ series. However, the rate at which stabilisation occurs is twice as fast for the $D_k$ series as for the $A_k$ series: the increment in $a_{\frac 1 4 + n, \frac 1 2}$ from $D_k$  to  $D_{k+1}$ stabilises when $2k - 5 \geq n$.

\paragraph{}
Finally, we turn to the $E_k$ singularities:

\begin{center}
\begin{tabular}{c|cc|ccccccccc}
&$ a_{\frac 1 4,0} $ &  $a_{\frac 1 4, \frac 1 2}$ & $a_{\frac 5 4, \frac 1 2}$ & $a_{\frac 9 4, \frac 1 2}$ & $a_{\frac {13} 4, \frac 1 2}$ & $a_{\frac {17} 4, \frac 1 2}$ & $a_{\frac {21} 4, \frac 1 2}$ & $a_{\frac {25} 4, \frac 1 2}$ & $a_{\frac {29} 4, \frac 1 2}$ & $a_{\frac {33} 4, \frac 1 2}$ & $ \dots $ \\
\hline
$E_6$ & 6 & 0 & 26 & 134 & 446 & 1320 & 3362 & 8068 & 17886 & 38010 & $ \dots $\\
$E_7$ & 7 & 0 & 30 & 154 & 512 & 1514 & 3854 & 9252 & 20510 & 43586 & $ \dots $\\
$E_8$ & 8 & 0 & 34 & 174 & 578 & 1708 & 4346 & 10430 & 23116 & 49124 & $ \dots $\\
\end{tabular}
\end{center}

\paragraph{}
The differences between successive $E_k$  singularities are shown below, and can be seen to stabilise to the same sequence as before, and the stabilisation happens twice as fast as compared to the $A_k$ series.

\begin{center}
\begin{tabular}{c|cc|ccccccccc}
&$ a_{\frac 1 4,0} $ &  $a_{\frac 1 4, \frac 1 2}$ & $a_{\frac 5 4, \frac 1 2}$ & $a_{\frac 9 4, \frac 1 2}$ & $a_{\frac {13} 4, \frac 1 2}$ & $a_{\frac {17} 4, \frac 1 2}$ & $a_{\frac {21} 4, \frac 1 2}$ & $a_{\frac {25} 4, \frac 1 2}$ & $a_{\frac {29} 4, \frac 1 2}$ & $a_{\frac {33} 4, \frac 1 2}$ & $ \dots $\\
\hline
$E_6 \to E_7$ & 1 & 0 & \textbf{4} & \textbf{20} & \textbf{66} & \textbf{194} & \textbf{492} & 1184  & 2624 & 5576 & $ \dots $\\
$E_7 \to E_8$ & 1 & 0 & \textbf{4} & \textbf{20} & \textbf{66} & \textbf{194} & \textbf{492} & \textbf{1178} & \textbf{2606} & 5538 & $ \dots $\\

\end{tabular}
\end{center}

\paragraph{}
In some sense, these patterns can be viewed as a quarter-BPS generalisation of the stringy McKay correspondence. It is intuitive that the multiplicities of quarter-BPS multiplets should increase every time an additional  $\mathbf P^1$ is added to the exceptional divisor, as is the case for half-BPS multiplets -- but why the regular increment only applies to multiplets with sufficiently low conformal dimension relative to the number of exceptional $\mathbf P^1$'s, and why the growth is governed by this particular sequence of numbers, is mysterious.

\paragraph{}
So far, we have only presented the character decompositions summed over all twisted sectors for all non-identity elements in $G$. It is possible to split these character decompositions further into contributions from twisted sectors for individual conjugacy classes in $G$. For any given conjugacy class, the sum over contributions from  sectors twisted by  group elements within the chosen conjugacy class does indeed admit an  integer-valued character decomposition. Furthermore, each conjugacy class can be associated in a natural way to a node in the Dynkin diagram, and it can be verified that the contribution from a given conjugacy class only depends, in a certain sense, on the environment that its associated node occupies within the Dynkin diagram. (Details are again provided in the Appendix.)

\begin{itemize}
\item For an $A_k$ singularity, the character contribution from the node in the $p^{\rm th}$ position from the left is equal to the contribution from the node  in the $p^{\rm th}$ position from the right.  (If $k$ is odd, with $k = 2l +1$, then one may view the diagram as a central node with two tails of length $l$ attached; this facilitates comparison with the $D$ and $E$ type singularities.) Moving from the central node(s) to the outer nodes, the $a_{\frac 1 4 + n , \frac 1 2 }$ coefficients always increase monotonically.

\item The Dynkin diagram for a $D_k$ singularity consists of a central node, attached to three tails of length $1$, $1$ and $k-3$ respectively. The Dynkin diagram for an $E_k$ singularity consists of a central node, attached to three tails of length $1$, $2$ and $k-4$. For any $D_k$ or $E_k$ singularity, the character contribution from a node situated on a tail of length $l$ in the $p^{\rm th}$ position from the end of the tail is equal to the character contribution in the $A_{2l+1}$ singularity from a node situated in the $p^{\rm th}$ position from either end of the $A_{2l+1}$  diagram. The contributions from the central nodes in the $D_k$ and $E_k$ diagrams are however distinct from contributions from nodes in $A$ type diagrams. Moving from the central node towards the outer-most node along any given branch in a $D$ or $E$ type diagram, the $a_{\frac 1 4 + n, \frac 1 2}$ coefficients always increase monotonically.
\end{itemize}

\paragraph{}
(In the Appendix, we illustrate these patterns in detail for the $A_3$ and $D_4$ examples.) Elliptic genera, unlike Euler characteristics, do not obey an inclusion-exclusion principle,\footnote{The inclusion-exclusion principle for Euler characteristics states that if $X$ is a complex algebraic variety and $Y$ is a closed subvariety, then $\chi(X) = \chi(Y) + \chi(X \backslash Y)$; it immediately follows that Euler characteristic of an algebraic variety increases by one when a point is replaced by a $\mathbf P^1$. There is no corresponding result for elliptic genera.} so it is rather surprising that there is a natural way to associate  elliptic genus contributions from individual conjugacy classes to particular irreducible blow-up divisors, and that these contributions appear to carry information about how the associated irreducible divisors are positioned with respect to the other irreducible divisors.

\section{ADE singularities in other orbifolds}

\paragraph{}
Although we have focused on the $\mathbf C^2 / G$ orbifold in our discussion so far, the quantities we have studied are also relevant to general orbifolds with singularities of ADE type. Our  $Z_{\mathbf C^2 / G,{\rm twisted}}$ quantities can therefore be viewed as a generic set of building blocks for the elliptic genera of all orbifold CFTs where the orbifolding procedure introduces ADE type singularities. Indeed, this was the author's original motivation for studying these quantities. In this section, we will develop the story in this more general setting, and we will explain how our $Z_{\mathbf C^2 / G,{\rm twisted}}$ quantities, whose numerical properties were studied in the previous section, are related to indices associated to sigma models with compact target space.

\paragraph{}
Let us therefore consider a two-dimensional $\mathcal N = (4,4)$ sigma model with target space $M$, where $M$ is now a general smooth hyperk\"ahler manifold of complex dimension two. Suppose that $M$ has an isometry $G$ that preserves the hyperk\"ahler structure, and suppose that the action of every non-identity element in $G$ has only isolated fixed points on $M$, which descend to ADE type singularities under the quotienting projection $M \to M / G$. Then the elliptic genus of the $M / G$ orbifold theory is
\begin{eqnarray*}
Z_{M / G}(q, \bar q, y)  = Z_M^G (q, \bar q, y)+ \sum_{x  \in  {\rm Sing}(M / G)} Z_{x, {\rm twisted}} (q,  y),
\end{eqnarray*}
where the untwisted sector contribution $Z_M^G(q, \bar q, y)$ is the index computed from $G$-invariant states on $M$, and ${\rm Sing}(M/G)$ denotes the singular locus of $M / G$.

\paragraph{}
Of course, the untwisted sector contribution $Z_M^G$ depends considerably on the global structure of the hyperk\"ahler manifold $M$. However, each twisted sector contribution  $ Z_{x, {\rm twisted}}$ only depends on the ADE classification of the singularity at $x$. \emph{Therefore, each $Z_{x , {\rm twisted}}$ is identical to the $Z_{\mathbf C^2 / G, {\rm twisted}}$ of corresponding ADE type.} In this way, our $Z_{\mathbf C^2 / G, {\rm twisted}}$ quantities are universal properties of ADE orbifolds.

\paragraph{}
(This conclusion follows, for instance, from the general orbifold formula for elliptic genera in \cite{anatoly1,anatoly1a, anatoly2, open1}, where the twisted sector contributions for isolated $x \in {\rm Sing}(M/G)$ are given solely in terms of the eigenvalues of the action of elements of ${\rm Stab}(x) \subset G$ on the tangent space $T_xM$. The qualitative difference between the untwisted and twisted sectors relies crucially on the fact that the identity element in $G$ fixes the whole of $M$, whereas the non-identity elements in $G$ have isolated fixed points.)

\paragraph{}
Now, consider the case where $M$ is compact, which implies that $M$ is either a  torus $T^4 = \mathbf C^2 / \Lambda$ or a K3 surface, and take $G$ to be a finite group of symplectic automorphisms, acting on $M$ with isolated fixed points that descend to du Val singularities on $M / G$. Then the minimal resolution $\widehat{M / G}$ is a smooth K3 surface\footnote{A symplectic automorphism $g \in {\rm Aut}(M)$ is one that fixes the generator $\Omega \in H^{2,0}(M)$. Given a finite group of symplectic automorphisms $G$, one can always find a K\"ahler form that is invariant under the action of $G$: indeed, if $\omega \in H^{1,1}(M)$ is any K\"ahler form, then the group average $\frac 1 {|G|} \sum_{g \in G} g^\star \omega$ is invariant under the action of $G$. Since a symplectic automorphism fixes the global non-vanishing holomorphic two-form $\Omega$, the canonical bundle is trivial on the open set of smooth points in $M / G$, and since minimal resolutions of ADE singularities are crepant resolutions, $\widehat{M/G}$ also has a trivial canonical bundle; see \cite{galoiscover} for example.}.

\paragraph{}
These orbifold K3 sigma models occur at very special points in the K3 moduli space. Such points in moduli space are characterised by the fact that the classes in $H^2({\rm K3})$ with self-intersection $-2$ that correspond to the exceptional rational curves are orthogonal to the generator $\Omega $ of $H^{2,0}({\rm K3})$ and also to the K\"ahler form $\omega$ in $ H^{1,1}({\rm K3});$\footnote{The position in moduli space is uniquely specified by how the real four-plane spanned by ${\rm Re} \ \Omega$, ${\rm Im} \ \Omega$, ${\rm Re}\exp(B + i\omega)$ and ${\rm Im}\exp(B + i\omega)$ embeds into $H^\star({\rm K3})\otimes \mathbf R$  modulo lattice automorphisms (where the lattice $H^\star({\rm K3})$ is endowed with the Mukai intersection form) \cite{symmetries1, symmetries2, symmetries3} (see also \cite{symmetries,huybrechts}). This is a physical version of the global Torelli theorem.} in particular, the Picard rank of the $K3$ is at least as large as the number of exceptional rational curves in the minimal resolution. (However, the exceptional $(-2)$-cycles carry non-trivial B-field flux \cite{wendland, aspinwallbfield}.)\footnote{In fact, there is no reason to expect that every K3 sigma model with ADE type singularities in the target space can be realised as an orbifold of $T^4$, or of another K3 surface. Nonetheless, even in these cases, the singularities will locally be indistinguishable from  the singularities that arise from orbifolding, provided that the B-field fluxes through the ADE rational curves are tuned to match the values obtained by the orbifolding procedure. Therefore, it is reasonable to believe that these sigma models will also contain sectors of states that wind around the singularities whose elliptic genus contributions are identical to the contributions from twisted sectors in orbifolds given by the $Z_{\mathbb C^2 / G, {\rm twisted}}$ functions.} These orbifold sigma models  have marginal operators  generating deformations equivalent to resolving the singularities \cite{ade4,ade1}, which can be viewed as deforming the K\"ahler form $\omega$ so that it is no longer orthogonal to the exceptional curves.

\paragraph{}
Since elliptic genera on compact spaces are always invariant under continuous deformations \cite{witteneg}, the elliptic genus of the orbifold, $Z_{M / G}(q,y)$,  is equal to the elliptic genus of the minimal resolution, $Z_{\widehat{M / G}}(q,y)$ (see also \cite{anatoly2} for a mathematical proof). Meanwhile, the quantity $Z_M^G(q,y)$ can be decomposed as a group average of twined elliptic genera $Z_M^g(q,y)$ for the manifold $M$,
\begin{eqnarray*}
Z_M^G (q,y)= \frac 1 {|G|} \sum_{g \in G} Z_M^g (q,y).
\end{eqnarray*}

\paragraph{}
For any compact complex manifold $X$ with trivial canonical bundle, the elliptic genus,  and more generally, the twined elliptic genera, can be defined geometrically as follows  \cite{gritzenko,creutzer}: if $g \in {\rm Aut}(X)$ is an automorphism fixing the generator of $H^{{\rm dim}_{\mathbb C}(X),0}(X)$, then the twined elliptic genus $Z_X^g(q,y)$ is given by a signed character for the action of $g^\star$ on the sheaf cohomology groups of $E_{q, y}$,
\begin{eqnarray*}
Z_X^g (q,y) = \sum_{i = 0}^{\dim_{\mathbf C} (X)} (-1)^i {\rm \ tr \ } g^\star |_{H^i(E_{q,y})},
\end{eqnarray*}
where $E_{q,y}$ is the formal power series of vector bundles,
\begin{eqnarray*}
E_{q,y} = y^{{\rm dim}_{\mathbf C}( X)  / 2} \wedge_{-y^{-1}}T_X^\star \bigotimes_{n \geq 1} \wedge_{-y^{-1}q^n}T_X^\star \bigotimes_{n \geq 1} \wedge_{-yq^n} T_X \bigotimes_{n \geq 0} {\rm Sym}_{q^n} (T_X \oplus T_X^\star).
\end{eqnarray*}
When $g$ is taken to be the identity automorphism, this reduces to the standard elliptic genus $Z_X(q,y)$ of $X$. By considering the Chern roots of $T_X$, it can be shown that $Z_X(q,y)$ is a weak Jacobi form of weight zero and index ${\rm dim}_{\mathbf C}( X ) / 2$ \cite{gritzenko}; in particular, it is independent of the $\bar q$ variable. Other twining genera obey similar modularity properties for congruence subgroups. The $\mathcal N = 4$ character decomposition of $Z_X(q,y)$ may then be  interpreted as the coefficients of a mock modular form associated to the weak Jacobi form $Z_X(q,y)$ \cite{zagier}.

\paragraph{}
If $M$ is chosen to be a K3 surface, then we are in a situation where both $M$ and $\widehat{M/G}$ are K3 surfaces. Because the moduli space of K3 sigma models is connected and the elliptic genus is invariant under continuous deformations, all K3 sigma models have the same elliptic genus. Furthermore,  finite-order symplectic automorphisms of K3 surfaces have been classified \cite{mukai,kondo}: they all give rise to ADE type singularities under quotienting, and the number of fixed points, as well as the eigenvalues of the action of the automorphisms on the tangent spaces at the fixed points, are uniquely determined by the orders of the automorphisms in the group $G$.  For any $g$ other than the identity element, the fixed points of $g$ are isolated, so the traces ${\rm  tr \ } g^\star |_{H^i(E_{q,y})}$ in the definition of $Z_X^g(q,y)$ can be written down directly in terms of the eigenvalues of the action of $g$ on the tangent spaces at the fixed points using the holomorphic Lefschetz fixed point formula \cite{atiyah}. Therefore, the twining genera are also uniquely determined by the order of the group element $g$.

\paragraph{}
The character decompositions of $Z_{{\rm K3}}^g$ have been computed in \cite{moonshine2, moonshine3}, and are tabulated below. The numbers in the left-hand column denote the order of the relevant automorphism $g$. The letters refer to the particular conjugacy class of $M_{24}$ associated by Mathieu moonshine to geometric automorphisms of the given order.\footnote{Mathieu moonshine states that for each $n $, there exists a non-trivial $M_{24}$ representation such that for each $g$, the coefficient $a_{\frac 1 4 + n, \frac 1 2}^g$ is equal to the character of this $M_{24}$ representation evaluated for the $M_{24}$ conjugacy class associated to $g$.}

\begin{center}
\begin{tabular}{c|cc|ccccccccc}
 & $ a^g_{\frac 1 4,0} $ &  $a^g_{\frac 1 4, \frac 1 2}$ & $a^g_{\frac 5 4, \frac 1 2}$ & $a^g_{\frac 9 4, \frac 1 2}$ & $a^g_{\frac {13} 4, \frac 1 2}$ & $a^g_{\frac {17} 4, \frac 1 2}$ & $a^g_{\frac {21} 4, \frac 1 2}$ & $a^g_{\frac {25} 4, \frac 1 2}$ & $a^g_{\frac {29} 4, \frac 1 2}$ & $a^g_{\frac {33} 4, \frac 1 2}$ & $ \dots $\\
\hline
1A & 20 & $-2$ & 90 & 462 & 1540 & 4554 & 11592 & 27830 & 61686 & 131100 & $ \dots $\\
2A & 4 & $-2$ & $-6$ & 14 & $-28$ & 42 & $-56$ & 86 & $-138$ & 188 & $ \dots $ \\
3A & 2 & $-2$ & 0 & $-6$ & 10 & 0 & $-18$ & 20 & 0 & $-30$ & $ \dots $\\
4B & 0 & $-2$ & 2 & $-2$ & $-4$ & 2 & 8 & $-2$  & $-10$ & 4 & $ \dots $\\
5A & 0 & $-2$ & 0 & 2 & 0 & $-6$ & 2 & 0 & 6 & 0 & $ \dots $\\
6A & $-2$ & $-2$ & 0 & 2 & 2 & 0 & $-2$ & $-4$ & 0 & 2 & $ \dots $\\
7AB & $-1$ & $-2$ & $-1$ & 0 & 0 & 4 & 0 & $-2$ & 2 & $-3$ & $ \dots $\\
8A & $-2$ & $-2$ & $-2$ & $-2$ & 0 & $-2$ & 0 & 2 & $-2$ & 0 & $ \dots $
\end{tabular}
\end{center}

\paragraph{}
Thus, in the case where $M$ is a  K3 surface, the relationship
\begin{eqnarray*}
Z_{\widehat{M / G}}  (q,y) = \frac 1 {|G|} \sum_{g \in G} Z_M^g (q,y) + \sum_{x \in {\rm Sing}(M / G)} Z_{x, {\rm twisted}} (q,y) 
\end{eqnarray*}
leads to a collection of relationships between the Mathieu twining genera and the quantities $Z_{\mathbf C^2 / G, {\rm twisted}}$ whose  properties were described in the previous section.

\paragraph{}
For instance, suppose $G$ is generated by a single symplectic automorphism of order 2. By the classification in \cite{mukai} (see also Table 1 in \cite{creutzer} where these results are collected), $M / G$ has eight $A_1$ singularities. Thus we obtain the relationship,
\begin{eqnarray*}
Z_{\rm K3} =  \frac 1 2 ( Z_{\rm K3} + Z_{\rm K3}^{2A} ) + 8 Z_{A_1, {\rm twisted}}.
\end{eqnarray*}
This relationship can be verified from the $a_{h,j}$ coefficients that we computed earlier for the $A_1$ singularity. For higher order automorphisms, one obtains similar relationships involving $Z_{A_k, {\rm twisted}}$ for $k \in \{1, 2, \dots, 7 \}$, which can be verified in the same way, and which enable one to solve for $Z_{A_k, {\rm twisted}}$ for $k \in \{1, 2, \dots, 7 \}$ in terms of the Mathieu twining functions.
\begin{eqnarray*}
Z_{\rm K3} & = & \frac 1 3 (Z_{\rm K3} + 2 Z_{\rm K3}^{3A}) + 6 Z_{A_2, {\rm twisted}}, \\
Z_{\rm K3} & = & \frac 1 4 (Z_{\rm K3} +  Z_{\rm K3}^{2A} + 2 Z_{\rm K3}^{4B} )+ 4 Z_{A_3, {\rm twisted}} + 2 Z_{A_1, {\rm twisted}}, \\
Z_{\rm K3} & = & \frac 1 5 (Z_{\rm K3} + 4 Z_{\rm K3}^{5A} ) + 4 Z_{A_4, {\rm twisted}}, \\
Z_{\rm K3} & = & \frac 1 6 (Z_{\rm K3} + Z_{\rm K3}^{2A} + 2 Z_{\rm K3}^{3A} + 2 Z_{\rm K3}^{6A}) + 2Z_{A_5, {\rm twisted}}+2Z_{A_2, {\rm twisted}} + 2Z_{A_1, {\rm twisted}} \\
Z_{\rm K3} & = & \frac 1 7 (Z_{\rm K3} + 6 Z_{\rm K3}^{7AB} ) + 3 Z_{A_6, {\rm twisted}},\\
Z_{\rm K3} & = & \frac 1 8 (Z_{\rm K3} + Z_{\rm K3}^{2A} + 2Z_{\rm K3}^{4B} + 4 Z_{\rm K3}^{8A} ) + 2 Z_{A_7, {\rm twisted}} + Z_{A_3, {\rm twisted}} + Z_{A_1, {\rm twisted}}.
\end{eqnarray*}

\paragraph{}
There are also examples of K3 surfaces whose groups of finite-order symplectic automorphisms are generated by more than one element. Their quotients contain combinations of du Val singularities of type $A_1, \dots , A_7$, $D_4, D_5, D_6$ and $E_6$. There are eighty-one cases altogether, and they are classified in \cite{galoiscover}. Each such example leads to a relationship between the respective Mathieu twining functions and the appropriate $Z_{\mathbf C^2 / G, {\rm twisted}}$ in the same way as before. To give just one example, the thirty-seventh case listed in \cite{galoiscover} is a $K3$ surface with a sympletic automorphism group isomorphic to the binary tetrahedral group (a group with one element of order $1$, one element of order $2$, eight elements each of orders $3$ and $6$, and six elements of order $4$), and the quotient $M / G$ contains two $A_2$ singularities, plus singularites of type $A_5$, $D_4$ and $E_6$. This leads to the relationship,
\begin{multline*}
Z_{\rm K3} = \frac 1 {24} (Z_{\rm K3} + Z_{\rm K3}^{2A} + 8 Z_{\rm K3}^{3A} + 6 Z_{\rm K3}^{4B} + 8Z_{\rm K3}^{6A}) \\  + Z_{E_6, {\rm twisted}} + Z_{D_4, {\rm twisted}} + Z_{A_5, {\rm twisted}} + 2 Z_{A_2, {\rm twisted}}.
\end{multline*}
Of course, the logic can be turned the other way round: the requirement for the existence of a relationship of this form provides a simple, non-trivial, consistency check for the existence of an $M$ with a symplectic automorphism group $G$ with given fixed point structure.

\paragraph{}
Instead of taking $M$ to be a K3 surface, one could also take $M$ to be a torus $T^4 = \mathbf C^2  / \Lambda$. As before, $G$ is a non-trivial group of discrete symplectic automorphisms with isolated fixed points. The resolved quotient $\widehat{M/G}$ is again a K3 surface. These complex tori have been classified in \cite{wendland}: there are eight cases, giving rise to various combinations of $A_1, \dots A_5$, $D_4$, $D_5$ and $E_6$ singularities in the quotient $M / G$. Again, one obtains relationships between twining genera and the twisted sector contributions from the singularities. For instance, the simplest case is where $G$ is the order two group generated by the Kummer involution $\iota$ sending $(x,y) \in \mathbf C^2 / \Lambda$ to $(-x,-y)$. Since the quotient $M / G$ has sixteen $A_1$ singularities, we obtain the relationship,
\begin{eqnarray*}
Z_{\rm K3} = \frac 1 2 (Z_{T^4} + Z_{T^4}^{\iota}) + 16 Z_{A_1, {\rm twisted}}.
\end{eqnarray*}
However, $T^4$ has trivial tangent bundle, and since the holomorphic Euler characteristic of a trivial bundle on $T^4$ is zero by Hirzebruch-Riemann-Roch, we have $Z_{T^4} = 0$. Furthermore, the Kummer involution $\iota$ has sixteen isolated fixed points whereas an order two symplectic automorphism of a K3 surface only has eight isolated fixed points \cite{creutzer,mukai}, so by the holomorphic Lefschetz fixed point formula, $Z_{T^4}^{\iota} = 2 Z_{\rm K3}^{2A}$. Thus the relationship reduces to $Z_{\rm K3} =  Z_{\rm K3}^{2A} + 16 Z_{A_1, {\rm twisted}}$, which is a rearrangement of a relationship that we have  already found. Similar relationships can be obtained for the other seven cases of $T^4 / G$ classified in \cite{wendland}.

\paragraph{}
The decomposition of $Z_{\widehat{M/G}}$ into a contribution from $G$-invariant quarter-BPS multiplets on $M$ plus contributions from ADE twisted sectors is useful for making sense of a puzzle in \cite{arnav1, arnav2}, for certain special cases. In \cite{arnav1, arnav2}, the authors study the \emph{unsigned} counts of quarter-BPS multiplets in K3 sigma models. In a generic K3 sigma model, the multiplicities of quarter-BPS multiplets are simply given by $ c_{(\frac 1 4 + n,\frac 1 2), (\frac 1 4, 0)}  = a_{(\frac 1 4 + n ,j)}$ and $c_{(\frac 1 4 + n,\frac 1 2), (\frac 1 4 , \frac 1 2 )} = 0$ for $n = 1,2, 3, \dots$, where $a_{(\frac 1 4 + n ,j)}$ are the dimensions of $M_{24}$ representations listed in the table above. However, on certain closed subvarieties within the moduli space, the quarter-BPS spectrum may be enhanced; in other words, the values of $ c_{(\frac 1 4 + n ,\frac 1 2 ), (\frac 1 4, 0)}  $ and $c_{(\frac 1 4 + n, \frac 1 2), (\frac 1 4 , \frac 1 2 )}$ may increase (in such a way that the difference $c_{(\frac 1 4 + n, \frac 1 2), (\frac 1 4, 0)} - 2 c_{(\frac 1 4 + n , \frac 1 2), (\frac 1 4 , \frac 1 2 )}$ remains equal to $a_{(\frac 1 4 + n , \frac 1 2)} $), and the puzzle is to characterise these jumping loci in the moduli space of K3 sigma models. On the special loci in moduli space where the K3 sigma model is an orbifold of the form $T^4 / G$, we can describe qualitatively how the enhancement of the quarter-BPS spectrum occurs. First, we have contributions from each ADE twisted sector, which, as described in the previous section, are approximately proportional to the number of nodes in the corresponding ADE Dynkin diagram; by a simple free field calculation of a refined index called the Hodge elliptic index (defined in \cite{arnav1}), which essentially reduces to observing that the twisted sectors have no right-moving fermion zero modes, these twisted sector contributions to the quarter-BPS spectrum can be seen to give no enhancement to the quarter-BPS spectrum. Therefore, all of the enhancements to the quarter-BPS spectrum on $T^4 / G$ come from $G$-invariant enhancements to the quarter-BPS spectrum on $T^4$, which, as described in \cite{arnav2}, depend on the K\"ahler and complex structure of the $T^4$, and are particularly large at special points in the moduli space (for instance at points where the $T^4$ is isomorphic to a product of two elliptic curves with complex multiplication and where the K\"ahler structure obeys an analogous condition). An immediate consequence of this description is that, if $G$ is a hyperk\"ahler-preserving automorphism group of a particular $T^4$, and $H$ is a subgroup of $G$, then the enhancements in the quarter-BPS spectrum of $T^4 / H$ are larger than the enhancements in the quarter-BPS spectrum of $T^4 / G$.

\section{Concluding remarks}

\paragraph{}
The most immediate question arising from our discussion is whether the sequence,
\begin{eqnarray*}
 4, \ \ 20, \ \ 66, \ \ 194, \ \ 492, \ \ 1178, \ \ 2606, \ \ 5530 , \ \ \dots
\end{eqnarray*}
which controls the growth of the ADE twisted sector character decompositions, has a natural origin. For example, one might speculate that the repetition of this sequence is a sign that sufficiently low-lying quarter-BPS multiplets experience some enhanced symmetry algebra. Alternatively, the fact that the multiplicities approach a stable pattern could also be a subtle consequence of modular invariance.

\paragraph{}
One might also wish to investigate whether the apparent correspondence between conjugacy classes and  irreducible blow-up divisors can be formulated as a more general statement for higher-dimensional target spaces in such a way that the quarter-BPS spectra from twisted sectors for individual conjugacy classes encode positional information about their associated divisors. If so, it would then be natural to investigate the fate of the quarter-BPS states under the blow-up deformation; for instance, one might conjecture that the quarter-BPS states associated to a given conjugacy class  deform continuously into states localised around their associated irreducible divisor in the resolution.

\paragraph{}
Finally, it would  be interesting to further explore the relationship between the ADE twisted sector character decompositions and twining genera. Indeed, there are twenty-three cases of umbral moonshine, of which $M_{24}$ is only one instance, and these twenty-three cases of umbral moonshine admit an ADE classification of their own \cite{umbral1,umbral2}.

 \section*{Acknowledgements}
 The author would like to thank Nathan Benjamin, Arnav Tripathy, Roberto Volpato, Milind Shyani, Shamit Kachru and David Tong for stimulating discussions. The author is supported by Gonville and Caius College and the ERC Grant agreement STG 279943.

\appendix

\section{Appendix: computations and formulae}

\subsection*{Computation of elliptic genera}

\paragraph{}
Here, we provide expressions for the twisted sector contributions to the elliptic genus on $\mathbf C^2 / G$,
\begin{eqnarray*}Z_{\mathbf C^2 / G,{\rm twisted}}(q, y)  : = \frac 1 {|G|} \sum_{g \in G \backslash \{ e \} } \sum_{h \in G} Z_{\mathbf C^2,(g,h)} (q, y). \end{eqnarray*}

\paragraph{}
First of all, note that a group element $h$ sends the vacuum state in the twisted sector associated to the group element $g$ to the vacuum state in the twisted sector associated to the group element $hgh^{-1}$. Therefore, for a given $g \in G \backslash \{ e \}$, we only need to sum over contributions from $h $ in the centraliser of $g$.

\paragraph{}
Next, we denote the eigenvalues for the action of a group element $g \in G$ on the coordinates of $\mathbf C^2$ by $\alpha(g) = e^{2\pi i\lambda (g)}$ and $\alpha(g)^{-1} =   e^{- 2 \pi i \lambda(g)}$. For $h $ in the centraliser of $g$, the action of $h$ on the coordinates $\mathbf C^2$ can be diagonlised simultaneously with the action of $g$, and we denote the eigenvalues for the action of $h$  with respect to the same basis by $\alpha(h) = e^{2\pi i\lambda (h)}$ and $\alpha(h)^{-1} =  e^{- 2 \pi i \lambda(h)}$. The contribution $Z_{\mathbf C^2,(g,h)}(q, y)$ is then given by
\begin{multline*}
Z_{\mathbf C^2,(g,h) } (q,  y) = \\  \prod_{n=0}^\infty \frac{(1 - \alpha(h) q^{n + \lambda(g)} y)(1 - \alpha(h)^{-1} q^{n + 1 - \lambda(g)}y)(1 - \alpha(h) q^{n + \lambda(g)} y^{-1})(1-\alpha(h)^{-1} q^{n+1-\lambda(g)}y^{-1})}{(1-\alpha(h)q^{n+\lambda(g)})^2(1-\alpha(h)^{-1}q^{n+1-\lambda(g)})^2}.
\end{multline*}

\paragraph{}
This formula follows immediately from the geometric formula for the elliptic genus of orbifolds, Equation 1.5 in \cite{anatoly1}, in which the contributions from the twisted sector for a group element whose fixed points are isolated are written in a way that only depends on the eigenvalues of the action of the group element on the tangent spaces at those fixed points. In \cite{anatoly1,anatoly2}, it is shown that this formula obeys key consistency conditions, including modularity properties, the DMVV formula for symmetric products \cite{DMVV}, and invariance under crepant resolution of singularities. The formula in \cite{anatoly1} is valid for (but not restricted to) general orbifolds with du Val singularities, so our results  apply not just to $\mathbf C^2 / G$, but also for arbitrary orbifolds with ADE singularities.

\paragraph{}
In the case of $\mathbf C^2 / G$, the numerator of the formula can be recognised as a sum over fermionic oscillators, and the denominator can be recognised as a sum over bosonic oscillators (written as summed geometric series). A version of the elliptic genus for $A_k$ type singularities has previously been obtained in this form from a localisation computation \cite{sungjay}; there, the formula includes the untwisted sector, regularised by introducing additional chemical potentials.

\subsection*{Discrete subgroups of $SU(2)$}

\paragraph{}
We now provide details of the discrete groups $G \subset SU(2)$ used to construct the orbifolds $\mathbf C^2 / G$, paying particular attention to the centralisers of the group elements and the eigenvalues $e^{2\pi i \lambda(g)}$ and $e^{2\pi i \lambda(h)}$. See \cite{ade1, johnson} for further details.

\paragraph{}
For $m \geq 2$, the $A_{m-1}$ singularity is obtained by taking $G$ to be the cyclic group of order $m$, generated by the element
\begin{eqnarray*}
a = \left( \begin{array}{cc} e^{2\pi i / m} & 0 \\ 0 &  e^{-2\pi i / m}\end{array}\right).
\end{eqnarray*}
Of course, for any element $g = a^k $ with $k \in \{1, \dots, m - 1 \}$, we have $\lambda(g) = \frac k m$, and since the centraliser $C(g)$ is the full group, the list of $\lambda(h) $ for $h \in C(g)$ contains one copy of $\frac p m$ for each $p \in \mathbf Z_m$. 

\paragraph{}
For $m \geq 2$, the $D_{m + 2}$ singularity is obtained by taking $G$ to be the binary dihedral group of order $4m$, generated by the two elements,
\begin{eqnarray*}
a = \left( \begin{array}{cc} e^{\pi i / m} & 0 \\ 0 &  e^{-\pi i / m}\end{array}\right), \ \ \ \ \ b =  \left( \begin{array}{cc} 0 & -1 \\ 1 & 0 \end{array}\right).
\end{eqnarray*}
The full group is $\{ 1, a, \dots, a^{2m - 1}, b, ab, \dots, a^{2m-1}b \}$, and the group elements obey the relation $a^m = b^2 = -1$.

\begin{itemize}
\item The element $g = -1 $ has  $\lambda(g) = \frac 1 2$. The centraliser $C(g)$ is the entire group, and the list of $\lambda(h)$ for $h \in C(g)$ contains one copy of $\frac p {2m}$ for each $p \in \mathbf Z_{2m}$, plus a further $2m$ copies of $\frac 1 4$.
\item For any $k \in \mathbf Z_{2m} \backslash \{ 0, m \} $, the element $g = a^k$  has $\lambda(g) = \frac k {2m}$. The centraliser $C(g)$ contains all elements of the form $a^p$, and so the list of $\lambda(h)$ for $h \in C(g)$ contains one copy of $\frac p{2m}$ for each $p \in \mathbf Z_{2m}$.
\item For any $k \in \mathbf Z_{2m}$, the element $g = a^k b$ has $\lambda(g) = \frac 1 4$. We have $C(g) = \{ 1, a^m, a^kb, a^{m+k}b \}$, and the list of $\lambda(h)$ for $h \in C(g)$ contains one copy of $\frac p 4$ for each $p \in \mathbf Z_4$.
\end{itemize}

\paragraph{}
Now let $x,y, z$ be the quaternion generators,
\begin{eqnarray*}
x = \left( \begin{array}{cc} i & 0 \\  0 & -i \end{array}\right), \ \ \ \ \ \ y = \left( \begin{array}{cc} 0 & 1\\ -1& 0 \end{array}\right), \ \ \ \ \ \ z = \left( \begin{array}{cc} 0 & i \\ i & 0   \end{array}\right).
\end{eqnarray*}
The $E_6$ singularity arises when $G$ is the binary tetrahedral group. The elements of this order $24$ group are the eight quaternions $\{ \pm 1, \pm x, \pm y, \pm z \}$, combined with sixteen elements of the form $\frac 1 2 (\pm 1 \pm x \pm y \pm z )$.
\begin{itemize}
\item The element $g = -1$ has $\lambda(g) = \frac 1 2 $. It commutes with the entire group, and the list of $\lambda(h)$ for $h \in C(g)$ contains one copy of $0$, one copy of $\frac 1 2$, six copies of $\frac 1 4$, eight copies of $\frac 1 6$ and eight copies of $\frac 1 3$.
\item There are six elements $g$ of order four with $\lambda(g) = \frac 1 4$. For each such element, the  list of $\lambda(h)$ for $h \in C(g)$ contains one copy  of $\frac p 4$ for each $p \in \mathbf Z_4$.
\item There are eight elements $g$ of order six with $\lambda(g) = \frac 1 6$, and a further eight elements $g$ of order three with $\lambda(g) = \frac 1 3$. For any of these elements, the list of  $\lambda(h)$ for $h \in C(g)$ contains one copy of $\frac p 6$ for each $p \in \mathbf Z_6$.
\end{itemize}

\paragraph{}
The $E_7$ singularity arises when $G$ is the binary octohedral group. This order $48$ group contains the quaternions $\{ \pm 1, \pm x, \pm y, \pm z \}$,  all elements of the form $\frac 1 2 (\pm 1 \pm x \pm y \pm z )$, and all elements obtained from $\frac 1 {\sqrt 2}(\pm 1 \pm x)$ or $\frac 1 {\sqrt 2} (\pm y \pm z)$ by permuting the set $\{x, y,z \}$.
\begin{itemize}
\item The element $g = -1$ has $\lambda(g) = \frac 1 2$. It commutes with the entire group, and the list of $\lambda(h)$ for $h \in C(g)$ contains one copy of $0$, one copy of $\frac 1 2$, eighteen copies of $\frac 1 4$, eight copies each of $\frac 1 3$ and $\frac 16$ and six copies each of $\frac 1 8 $ and $\frac 3 8$.
\item There are twelve elements of order four with $\lambda(g) = \frac 1 4$, for which the list of $\lambda(h)$ for $h \in C(g)$  contains one copy of $\frac p 4$ for each $p \in \mathbf Z_4$.
\item There are eight elements of order six with $\lambda(g) = \frac 1 6$ and a further eight elements of order three with $\lambda(g) = \frac 1 3$. For any of these elements, the list of $\lambda(h)$ for $h \in C(g)$  contains  one copy of $\frac p 6$ for each $p \in \mathbf Z_6$.
\item There are six elements of $g$ order eight with $\lambda(g) = \frac 1 8$, plus a further six elements $g$ of order four with $\lambda(g) = \frac 1 4$,  plus six elements of order eight with $\lambda(g) = \frac 3 8$. For any of these elements, the list of $\lambda(h)$ for $h \in C(g)$ contains  one copy of $\frac p 8$ for each $p \in \mathbf Z_8$.
\end{itemize}

\paragraph{}
The $E_8$ singularity arises when $G$ is the binary octohedral group. This order $120$ group contains the quaternions $\{ \pm 1, \pm x, \pm y, \pm z \}$,  all elements of the form $\frac 1 2 (\pm 1 \pm x \pm y \pm z )$ and all elements obtained from $\frac 1 2 (0.1 \pm x \pm \varphi^{-1} y \pm \varphi z )$ by even permutations on the set $\{ 1, x, y, z \}$, where $\varphi = \frac 1 2 (1 + \sqrt{5})$.

\begin{itemize}
\item The element $g = -1$ has $\lambda(g) = \frac 1 2$. The list of $\lambda(h)$ for $h \in C(g)$ contains one copy of $0$, one copy of $\frac 1 2$, thirty copies of $\frac 1 4$, twenty copies each of $\frac 1 6 $ and $\frac 1 3$, and twelve copies each of $\frac 1 {10}, \frac 1 5, \frac 3 {10}$ and $\frac 2 5$.
\item There are thirty elements $g$ of order four with $\lambda(g) = \frac 1 4$, for which the list of $\lambda(h)$ for $h \in C(g)$ contains one copy of $\frac p 4$ for each $p \in \mathbf Z_4$.
\item There are twenty elements $g$ of order six with $\lambda(g) = \frac 1 6$, and a twenty elements $g$ of order three with $\lambda(g) = \frac 1 3$. For each such $g$, the list of $\lambda(h)$ with $h \in C(g)$ contains one copy of $\frac p 6$ for each $p \in \mathbf Z_6$.
\item There are twelve elements $g$ of order ten with $\lambda(g) = \frac 1 {10}$, plus another twelve of order five with $\lambda(g) = \frac 1 5$, plus another twelve of order ten with $\lambda(g) = \frac 3 {10}$, plus another twelve of order five with $\lambda(g) = \frac 2 5$. For each such $g$, the list of $\lambda(h)$ with $h \in C(g)$ contains one copy of $\frac p {10}$ for each $p \in \mathbf Z_{10}$.
\end{itemize}

\subsection*{Decomposition into characters}

\paragraph{}
The $\mathcal N = 4$ superconformal algebra at central charge $ c = 6$ has two short multiplets whose highest weight states have $L_0$ and $J_0$ eigenvalues $(h, j) = (\frac 1 4, 0) $ and $(h, j) = (\frac 1 4, \frac 1 2)$ respectively, and a sequence of long multiplets with $(h, j) = (\frac 1 4 + n, \frac 1 2)$ for $n = 1,2,3,\dots$ . The characters of these multiplets are computed in \cite{characters1, characters2}, and are reproduced below for convenience.
\begin{eqnarray*}
{\rm ch}_{\frac 1 4, 0}(q,y) & = & \prod_{m = 1}^\infty \frac{(1 - q^m y)^2(1 - q^{m-1}y^{-1})^2}{(1-q^m)^2(1-q^m y^2)(1- q^{m-1}y^{-2})} \times \sum_{p = -\infty}^{\infty} \frac{q^{2p^2}y^{4p}(q^p y + 1)}{q^p y - 1}, \\
{\rm ch}_{\frac 1 4, \frac 1 2}(q,y) & = & - 2{\rm ch}_{\frac 1 4, 0}(q,y)  - y \prod_{m = 1}^\infty \frac{(1 - q^m y)^2(1 - q^{m-1}y^{-1})^2}{(1-q^m)}, \\
{\rm ch}_{\frac 1 4 + n , \frac 1 2}(q,y) & = & - q^ny \prod_{m = 1}^\infty \frac{(1 - q^m y)^2(1 - q^{m-1}y^{-1})^2}{(1-q^m)}, \ \ \ \ \ \ \ \  \ \  \ \ \ n = 1,2,3, \dots
\end{eqnarray*}

\paragraph{}
Following \cite{zagier}, an efficient method for performing the decomposition is to note that $Z(q,1)$ is the Witten index, which is equal to $a_{\frac 1 4, 0} - 2 a_{\frac 1 4, \frac 1 2}$. Subtracting $Z(q,1)\times{\rm ch}_{\frac 1 4, 0}(q,y)$ from $Z(q,y)$, then dividing the remainder by ${\rm ch}_{\frac 1 4, \frac 1 2}(q,y)+ 2{\rm ch}_{\frac 1 4, 0}(q,y)$, leaves a power series whose $q^n$ coefficient (for $n = 0,1,2, \dots$) is $a_{\frac 1 4 + n, \frac 1 2}$.

\subsection*{Splitting the character decomposition into contributions from individual conjugacy classes}

\paragraph{}
As mentioned in the main text, the non-identity conjugacy classes in $G$ are in one-to-one correspondence with the nodes in the Dynkin diagram (or equivalently, to the exceptional rational curves in the minimal resolution of $\mathbf C^2 / G$). The Witten index contribution from the states in twisted sectors associated to all group elements in a single conjugacy class ${\rm Conj}(g')$ is $ \frac 1 {|G|} \sum_{g \in {\rm Conj}(g')} |C(g)|$, which is equal to one. We now discuss the quarter-BPS multiplicities $a_{\frac 1 4 + n, \frac 1 2,{\rm Conj}(g')}$ arising from any given conjugacy class ${\rm Conj}(g')$.

\paragraph{}
As will soon become apparent, it is possible to associate conjugacy classes ${\rm Conj}(g') \subset G$ in a natural way to nodes in the Dynkin diagram. In doing so, one finds that the quarter-BPS multiplicities $a_{\frac 1 4 + n, \frac 1 2,{\rm Conj}(g')}$ coming from a given conjugacy class ${\rm Conj}(g')$ only depend on the environment that its associated node occupies within the Dynkin diagram. This does not require explicit computation to verify; it follows immediately from coincidences between the eigenvalues $\lambda(g)$, and $\lambda(h)$ for $h \in C(g)$, for the various discrete subgroups $G$.

\paragraph{}
For instance, consider the cyclic group of order four, which corresponds to the $A_3$ singularity. This has three non-identity conjugacy classes: $\{ a \}, \{ a^2 \} $ and $\{ a^3 \}$. One may associate these conjugacy classes with the three nodes on the $A_3$ Dynkin diagram, in that order. From their eigenvalues, it is clear that the contributions from the twisted sector for $a$ and the contribution from the twisted sector for $a^3$ (i.e. the twisted sectors associated to the two outer nodes on the Dynkin diagram) are equal to one another, but are distinct from the contribution from the twisted sector for $a^2$ (i.e the twisted sector associated to the central node on the Dynkin diagram). For concreteness, we tabulate the individual contributions below:

\begin{center}
\begin{tabular}{c|cc|ccccccccc}
$A_3$ node: &$ a_{\frac 1 4,0} $ &  $a_{\frac 1 4, \frac 1 2}$ & $a_{\frac 5 4, \frac 1 2}$ & $a_{\frac 9 4, \frac 1 2}$ & $a_{\frac {13} 4, \frac 1 2}$ & $a_{\frac {17} 4, \frac 1 2}$ & $a_{\frac {21} 4, \frac 1 2}$ & $a_{\frac {25} 4, \frac 1 2}$ & $a_{\frac {29} 4, \frac 1 2}$ & $a_{\frac {33} 4, \frac 1 2}$ & $ \dots $\\
\hline
$\{ a \}$ & 1 & 0 & 6 & 28 & 98 & 282 & 728 & 1734 & 3864 & 8182 & $ \dots $\\
$\{ a^2 \}$ & 1 & 0 & 2 & 16 & 46 & 146 & 356 & 878 & 1916 & 4114 & $ \dots $ \\
$\{ a^3 \}$ & 1 & 0 & 6 & 28 & 98 & 282 & 728 & 1734 & 3864 & 8182 & $ \dots $\\

\end{tabular}
\end{center}

\paragraph{}
If we instead take the binary dihedral group of order eight, which is isomorphic to the quaternion group, we obtain the $D_4$ singularity. There are four non-identity conjugacy classes: $\{ a^2 \}$, $\{ a, a^3 \}$, $\{ b, a^2 b \}$ and $\{ab, a^3 b \}$, which can be rewritten as $\{ -1 \}$, $\{ x, -x \}$, $\{ y, - y\}$ and $\{z, -z \}$ to make the isomorphism with the quaternion group more apparent. One may associate the $\{ -1 \}$ conjugacy class to the central node in the $D_4$ Dynkin diagram, and then associate the $\{ x, -x \}$, $\{ y, -y \}$ and $\{ z, -z \}$ conjugacy classes to the three outer nodes. From their eigenvalues, and the eigenvalues of the elements in their centralisers, it is clear that the character contribution from the twisted sectors for $\{ x, -x\}$, for  $\{ y, -y \}$ and for $\{ z, -z \}$ each agree with the character contribution from $\{ a \}$, or from $ \{ a^3 \}$, for the $A_3$ singularity. This is reflected in how the three outer nodes in the $D_4$ Dynkin diagram are each one edge away from the central node, the same being true of the two outer nodes in the $A_3$ Dynkin diagram. However, the contribution from the twisted sector for the $\{ -1 \}$ conjugacy class in the binary dihedral group, which is associated with the central node in the $D_4$ Dynkin diagram, is distinct from all other contributions seen so far.

\begin{center}
\begin{tabular}{c|cc|ccccccccc}
$D_4$ node: &$ a_{\frac 1 4,0} $ &  $a_{\frac 1 4, \frac 1 2}$ & $a_{\frac 5 4, \frac 1 2}$ & $a_{\frac 9 4, \frac 1 2}$ & $a_{\frac {13} 4, \frac 1 2}$ & $a_{\frac {17} 4, \frac 1 2}$ & $a_{\frac {21} 4, \frac 1 2}$ & $a_{\frac {25} 4, \frac 1 2}$ & $a_{\frac {29} 4, \frac 1 2}$ & $a_{\frac {33} 4, \frac 1 2}$ & $ \dots $\\
\hline
$\{ -1 \}$ & 1 & 0 & 0 & 10 & 20 & 78 & 170 & 450 & 942 & 2080 & $ \dots $\\
$\{ x, -x\}$ & 1 & 0 & 6 & 28 & 98 & 282 & 728 & 1734 & 3864 & 8182 & $ \dots $\\
$\{ y, -y \}$ & 1 & 0 & 6 & 28 & 98 & 282 & 728 & 1734 & 3864 & 8182 & $ \dots $\\
$\{ z, -z \}$ & 1 & 0 & 6 & 28 & 98 & 282 & 728 & 1734 & 3864 & 8182 & $ \dots $\\

\end{tabular}
\end{center}

\paragraph{}
Similar patterns can be deduced for higher order $D$ and $E$ type singularities. The general pattern is as summarised in the main text. 

\paragraph{}
Finally, we examine how the character decomposition contribution from the $\{ a^p \}$ conjugacy class for $A_k$ grows with the number of nodes $k$ in the Dynkin diagram (with $p$ fixed). This  contribution is the contribution associated to a node situated in the $p^{\rm th}$ position from either end of the chain in the $A_k$ diagram, and if $k$ is odd, this is also the contribution from a node situated in the $p^{\rm th}$ position from the end of any tail of length $\frac 1 2 (k-1)$ in a $D$ or $E$ type diagram. By explicit computation, one observes that the growth of $a_{\frac 1 4  + n, \frac 1 2, {\rm Cl}(\{ a^p \})}$, as $k$ increases, eventually stabilises. In the table below, we give expressions for these character contributions, for $k$ sufficiently large that the growth has stabilised. We also give the corresponding expressions for the central node in the $D_k$ diagrams.

\begin{center}
\begin{tabular}{c|cc|cccccc}
$A_k$ node: &$ a_{\frac 1 4,0} $ &  $a_{\frac 1 4, \frac 1 2}$ & $a_{\frac 5 4, \frac 1 2}$ & $a_{\frac 9 4, \frac 1 2}$ & $a_{\frac {13} 4, \frac 1 2}$ & $a_{\frac {17} 4, \frac 1 2}$ & $a_{\frac {21} 4, \frac 1 2}$ & $ \dots $\\
\hline
($k \geq 1$) & & & ($k \geq 2$)  & ($k \geq 3$) & ($k \geq 4$) & ($k \geq 5$) & ($k \geq 6$) \\
$\{ a \}$ & 1 & 0 & $ 3 + k$ & $13 + 5k$ & $40+18k$ & $111+53k$ & $271 +139k$ & $ \dots $\\
\hline
($k \geq 3$) & & &  & ($k \geq 4$) & ($k \geq 5$) & ($k \geq 6$) & ($k \geq 7$) \\
$\{ a^2 \}$ & 1 & 0 & $ 2$ & $9+k$ & $25+3k$ & $69+11k$ & $160+28k$ & $ \dots $\\
\hline
($k \geq 5$) & & &  & & ($k \geq 6$) & ($k \geq 7$) & ($k \geq 8$) \\
$\{ a^3 \}$ & 1 & 0 & $ 2$ & $8$ & $23+k$ & $61+3k$ & $141+9k$ & $ \dots $\\
\hline
($k \geq 7$) & & &  & & & ($k \geq 8$) & ($k \geq 9$) \\
$\{ a^4 \}$ & 1 & 0 & $ 2$ & $8$ & $22$ & $59+k$ & $135+3k$ & $ \dots $\\
\hline
($k \geq 9$) & & &  & & & & ($k \geq 10$) \\
$\{ a^5 \}$ & 1 & 0 & $ 2$ & $8$ & $22$ & $58$ & $133+k$ & $ \dots $\\
\hline
($k \geq 11$) & & &  & & & &  \\
$\{ a^6 \}$ & 1 & 0 & $ 2$ & $8$ & $22$ & $58$ & $132$ & $ \dots $\\
\hline
$\vdots$ & $\vdots$ & $\vdots$ & $\vdots$ & $\vdots$ & $\vdots$ & $\vdots$ &  $\vdots$ \\
 & & & & & & & & \\
\hline
$D_k$ node: & & & ($k \geq 4$)  & ($k \geq 5$) & ($k \geq 6$) & ($k \geq 7$) & ($k \geq 8$) \\
$\{ -1 \}$ & 1 & 0 & 0 & 6 & 8 & 34 & 58 & $ \dots $\\
\end{tabular}
\end{center}

\paragraph{}
Remarkably, the growth pattern for the entire Dynkin diagram (described in the main text) always stabilises at a lower value of $k$ than the growth patterns for the individual nodes --  though of course, these growth patterns are consistent with one another.

\end{document}